\def\aa{{A\&A}}

\def\aj{{AJ}}
\def\annrev{{ARA\&A}}
\def\apj{{ApJ}}
\def\apjs{{ApJS}}

\def\mnras{{MNRAS}}

\documentclass[cup5b]{caps}
\usepackage{graphicx}
\usepackage{amssymb}
\usepackage{ociwsymp3}   

\HeadText{B. M. Poggianti}

\begin{document}

\pagenumbering{arabic}

\author[]{BIANCA M. POGGIANTI\\Astronomical Observatory, Padova, Italy}

\chapter{Modeling Stellar Populations in \\ Cluster Galaxies}

\begin{abstract}
In this review I highlight the role played by spectrophotometric
models in the study of galaxy evolution in distant clusters. I
summarize the main achievements of the modeling of k+a
spectra, the derivation of the star formation rate in emission-line
galaxies, and the stellar ages of red cluster galaxies. The current
knowledge of the dependence of the star formation histories on the 
galaxy morphology and luminosity is also presented.
\end{abstract}

\section{Introduction}
The goal of any spectrophotometric modeling is to reconstruct the 
star formation history of galaxies, to learn what has been the
star formation rate (SFR) at each epoch and, as a consequence, the enrichment
history of metals and the evolution of the galactic luminosity and mass.
In this review I focus on galaxies in mostly distant clusters ($z>0.2$),
and present the main results obtained using
spectrophotometric modeling as a tool to interpret the data.

Quite a few of the spectrophotometric models available today have the 
ability to include the three main components that produce the
spectral energy distribution of a galaxy: (1) the stellar contribution;
(2) the emission lines and emission continuum from gas ionized by young 
massive stars, and (3) the extinction due to dust. 
Nowadays it is common to include 
stellar populations of different metallicities, and models can have 
sufficient wavelength resolution ($3-4$ \AA \ or better) to study stellar
absorption features in detail.

In this context it is useful to think of three types of spectra: 
those {\it with emission lines} and, among those without emission lines,
{\it k+a} and {\it passive} spectra.
For each one of these types of spectra there is a stellar time scale
associated: $< 5 \times 10^7 $ yr for emission-line spectra (the lifetime
of the massive stars able to ionize the gas); between $5 \times 10^7 $
and $1.5 \times 10^9 $ yr for k+a spectra (the period during which
stars with strong Balmer lines 
dominate the integrated spectrum of a stellar system); 
and $> 1.5 \times 10^9 $ yr
for passive spectra. Each one of these spectral types is described and
discussed below.

Whenever my own work will be mentioned in the following, credit
should be given to my collaborators of the MORPHS group (H. Butcher,
W. Couch, A. Dressler, R. Ellis, A. Oemler, R. Sharples, I. Smail)
for distant cluster work, and of the Coma collaboration (T. Bridges,
D. Carter, B. Mobasher, Y. Komiyama, S. Okamura, N. Kashikawa, et al.) 
for results concerning the Coma cluster.

\begin{figure*}[t]
\centering
\includegraphics[width=0.90\columnwidth,angle=0,clip]{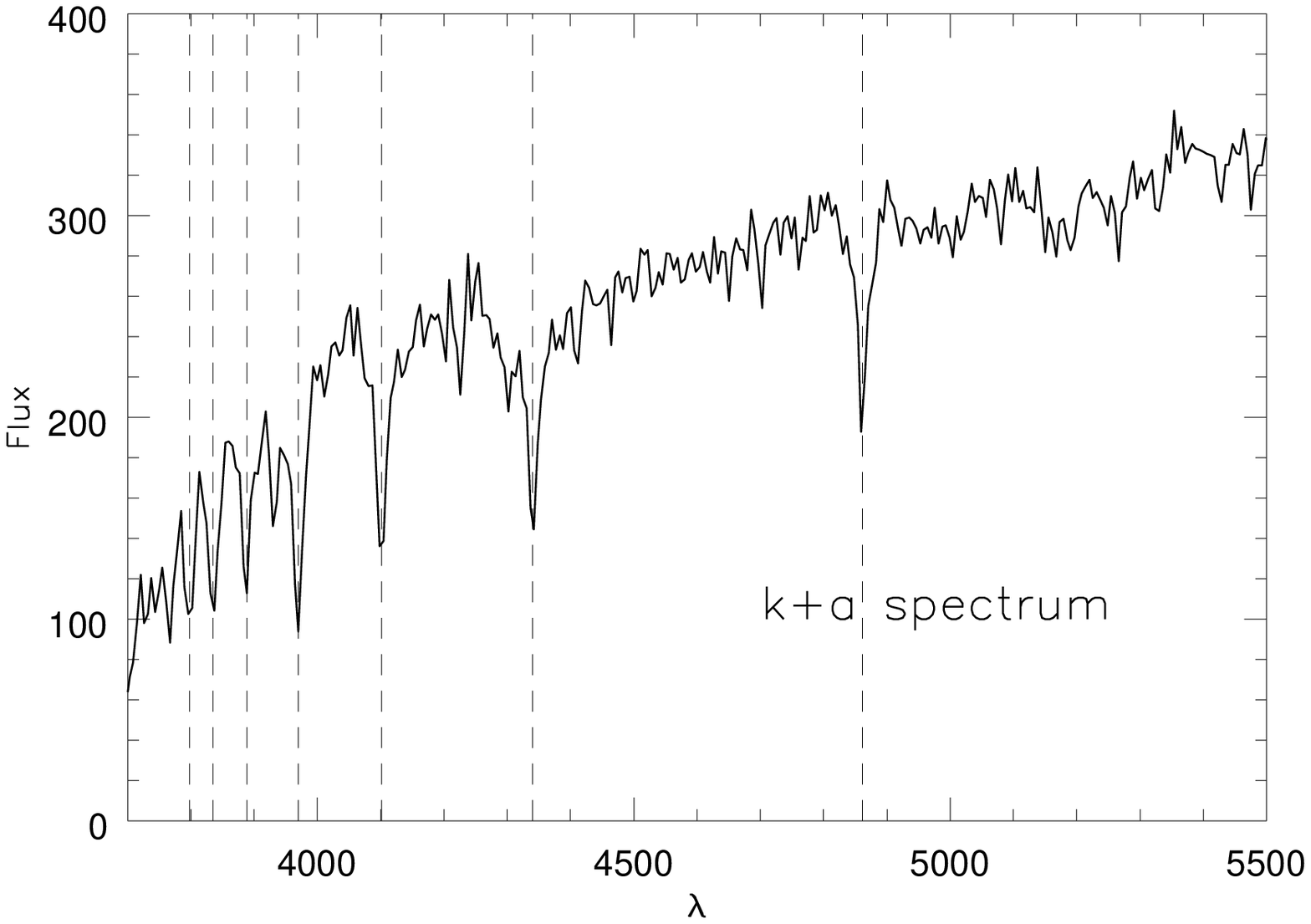}
\includegraphics[width=0.90\columnwidth,angle=0,clip]{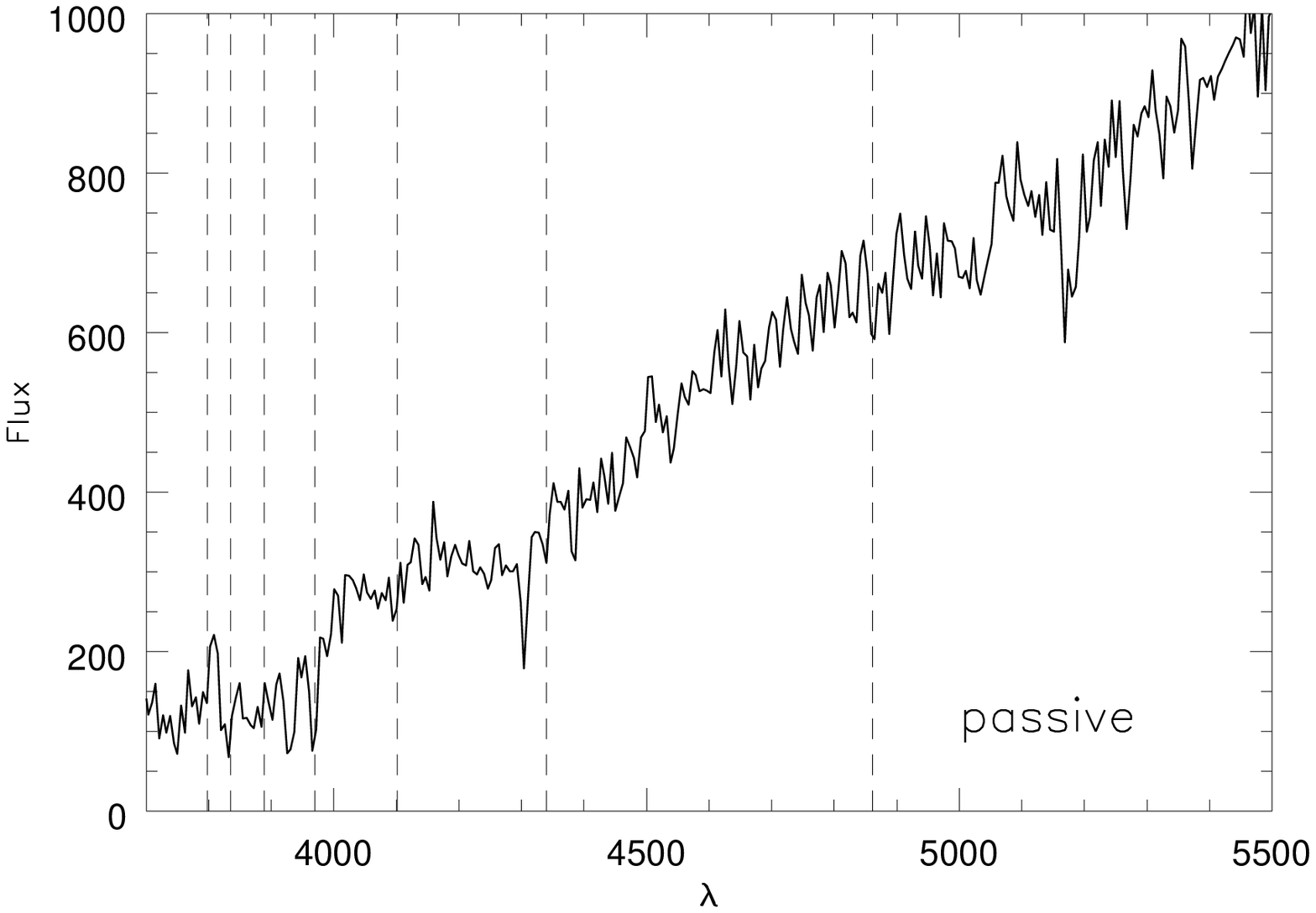}
\vskip 0pt \caption{
Example of a k+a spectrum ({\it top}) and of a passive spectrum 
({\it bottom}). The lines of the Balmer series (from left to right
$\rm H\theta$, $\rm H\eta$, $\rm H\zeta$, $\rm H\epsilon$, $\rm H\delta$,
$\rm H\gamma$, $\rm H\beta$) are highlighted by dashed lines.
\label{kacarnegie}}
\end{figure*}

\section{k+a Spectra}

Historically, k+a spectra (originally named ``E+A's'', also known
as $\rm H\delta$-strong galaxies, etc.) were the
first type of spectra to be noticed and modeled in distant cluster studies.
They are spectra with no emission lines and strong Balmer lines in
absorption [EW$(\rm H\delta) > 3$ \AA]. A strong-lined example of
a k+a galaxy is shown in Figure~\ref{kacarnegie}
and contrasted with a passive spectrum.

The second paper that appeared presenting spectra of distant cluster galaxies 
already pointed out what is still the current interpretation
of (the strongest) k+a spectra, that they ``indicate a large
burst of star formation $10^9$ years before the light left the
galaxy'' (Dressler \& Gunn 1983).  A large number of successive papers
have discussed k+a spectra in distant clusters (e.g., Henry \& Lavery 1987; 
Fabricant, McClintock, \& Bautz 1991; Fabricant, Bautz, \& McClintock 1994; 
Dressler \& Gunn 1992; Belloni et al. 1995; Belloni \& Roeser 1996; Fisher et 
al. 1998; Couch et al. 1998; Balogh et al. 1999; Dressler et al. 1999; 
Poggianti et al. 1999; Bartholomew et al. 2001; Ellingson et al. 2001; Tran et 
al. 2003); those that have performed spectrophotometric modeling to interpret 
them include Dressler \& Gunn (1983), Couch \& Sharples (1987), Newberry, 
Boroson, \& Kirshner (1990), Charlot \& Silk (1994), Jablonka \& Alloin (1995),
Abraham et al. (1996), Barger et al. (1996), Poggianti \& Barbaro (1996, 1997),
Morris et al. (1998), Bekki, Shioya, \& Couch (2001), Shioya,
Bekki, \& Couch (2001), and Shioya et al.  (2002).
The main conclusions of k+a modeling can be summarized as follows:

\begin{figure*}[t]
\includegraphics[width=1.00\columnwidth,angle=0,clip]{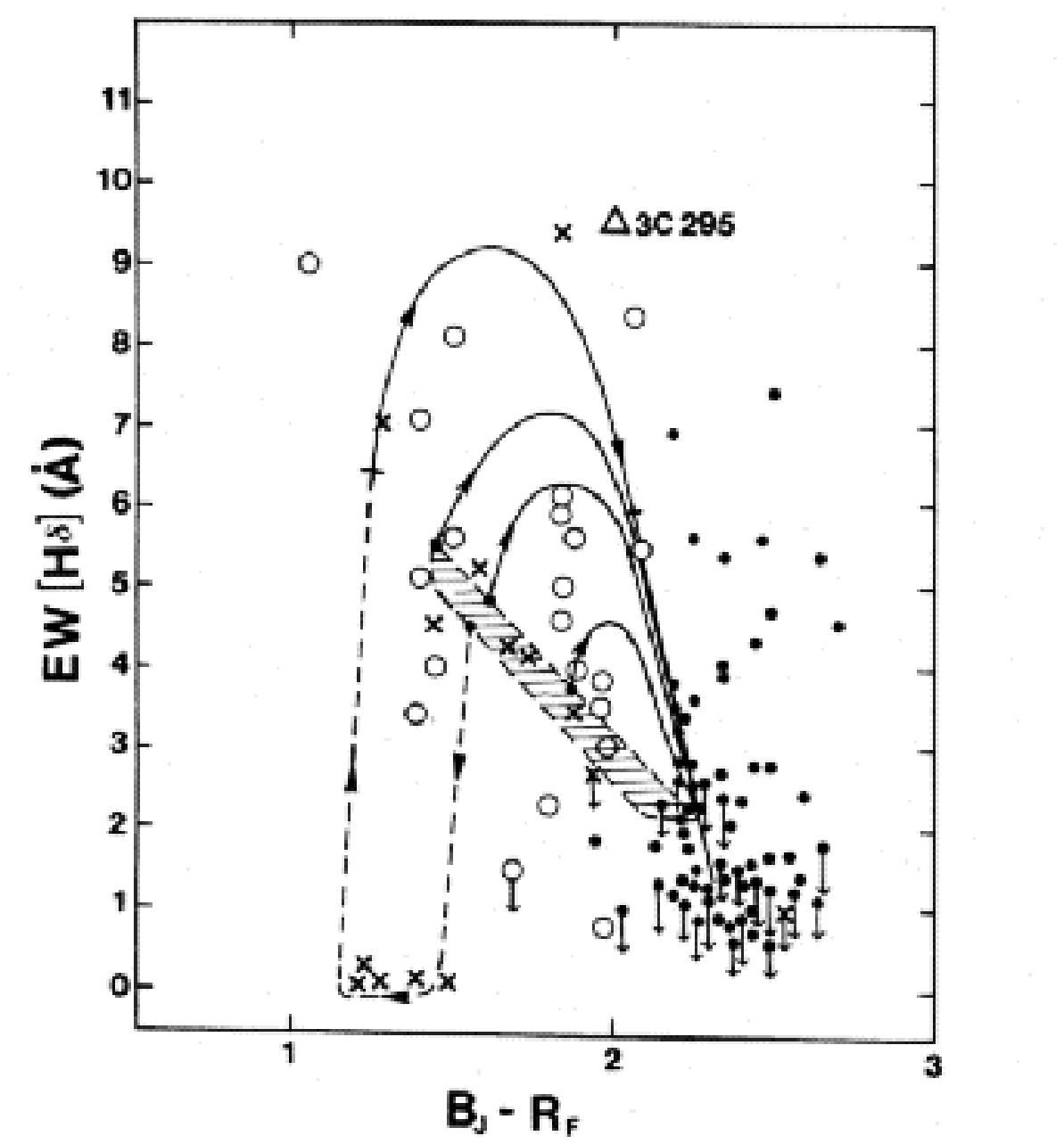}
\vskip 0pt \caption{
Color versus rest-frame equivalent width of $\rm H \delta$
of galaxies in three clusters at $z=0.3$ (Couch \& Sharples 1987).
Emission-line galaxies are indicated by crosses and non-emission-line
galaxies by filled (red galaxies) and empty (blue galaxies) circles.
The dashed region shows the locus of typical spirals today,
which generally display emission lines and therefore are comparable to the 
crosses in this diagram.
Ellipticals occupy the bottom right region of this
diagram, at $B-R>2$ mag and EW($\rm H\delta)<1$ \AA.
The lines are evolutionary sequences of galaxies in which star
formation is halted with or without a starburst preceding the
quenching of star formation. 
The former case is shown as a dotted+continuous line
leading downward from the spiral sequence, while examples
of truncated models with no starbursts are the solid tracks leading
upward from the spiral sequence.
\label{fig:cs3.ps}}
\end{figure*}

\begin{enumerate}
\item
In k+a galaxies, star formation stopped typically between $5 \times 10^7 $
and $1.5 \times 10^9 $ yr before the epoch at which the galaxy is observed.
The evolution of the equivalent width of $\rm H\delta$ versus color is shown 
in Figure~\ref{fig:cs3.ps}.  This diagram reproduces
a plot from the influential paper of Couch \& Sharples (1987).
Soon after the moment when star formation is interrupted, the galaxy moves 
in this diagram toward higher EW($\rm H\delta$) and progressively redder
colors, reaching a maximum in EW($\rm H\delta$) after $(3-5) \times 10^8 $ 
yr. Then, the strength of the line starts to decline until the 
locus occupied by typical ellipticals is reached, in the right bottom
corner of the plot.

\item
Spectra with very strong EW($\rm H\delta$) (above 5 \AA)
require a starburst prior to the truncation of the star formation, with
high galactic mass fractions involved in the burst (10\%--20\% or higher).
k+a spectra with a moderate $\rm H\delta$ line, on the other hand, 
may be post-starburst
galaxies in a late stage of evolution, but may also be reproduced
by simply interrupting ``normal'' star formation activity. The 5 \AA
$\,$ limit mentioned above should only be considered indicative:
given the strong dependence of the EW value on the
method adopted to measure it, the strength of the lines should 
always be measured in the same way on data and models.

\item
A combination of EW($\rm H\delta$) and color, such as that shown in 
Figure~\ref{fig:cs3.ps}, can
help in roughly estimating the time elapsed since the halting
of star formation. A more sophisticated method to age date the starburst
using Balmer-line indices can be found in Leonardi \& Rose (1996).

\item
A slowly declining SFR, with a time scale comparable
to, or longer than,
that of the k+a phase (1 Gyr or longer), is not able to produce a
k+a spectrum, which is more easily obtained if the star formation
is halted quickly. The reason for such behavior is
that while the SFR is slowly declining emission lines are still present
in the galaxy spectrum, thus the galaxy will not be classified as a k+a.
When finally the star formation terminates, the contribution of A-type stars
to the integrated spectrum is low, due to the low average level of star
formation during the previous Gyr.
\end{enumerate}

A great advantage of the k+a spectral classification is the fact
that, in principle, it requires a single measurement (the $\rm H\delta$ line),
though this must be of sufficiently high signal-to-noise ratio
and must be coupled with a complete inspection of the spectrum to
verify the absence of any emission line. Analysis of the other higher-order
lines of the Balmer series usually greatly helps in confirming
the strength of $\rm H\delta$.
Moreover, modeling of k+a spectra is relatively simple: 
the age-metallicity degeneracy that affects the interpretation of
passive spectra is not an issue here, because the effect in the
strength of the Balmer lines is so large that only age can be the cause 
for it. The strong Balmer lines arise in A- to F-type stars,
which are stars of $\sim 2$ solar masses on the main sequence, hence in
a well known and easily modeled evolutionary phase.

The importance of k+a modeling depends obviously on the importance
of the k+a phenomenon in cluster galaxy data. Many works (see list above)
have found distant
clusters to host a significant number of k+a galaxies. 
If k+a spectra occur proportionally
more in distant clusters than in the field at similar redshifts
(as found, e.g., by Dressler et al. 1999; but see 
Balogh et al. 1999 for a different
view), then they provide strong evidence that star formation is rapidly
quenched
in clusters, and they could be used as tracers both of the star formation
history of galaxies infalling into clusters and of the infall history
itself. The strongest k+a spectra are a solid signature of a
recent starburst; whether this burst was or was not induced
by the cluster environment is still an open issue and a subject of ongoing
investigation.


Finally, it is not a mere intellectual exercise to note that if a
galaxy formed all of its stars on a short time scale, as ellipticals
are expected to do in a monolithic-collapse
scenario of galaxy formation, then there must have been
a phase/epoch when it displayed a spectacular k+a spectrum 
of an almost pure single stellar population dominated by A stars.  In an 
$\Omega_{\Lambda}=0.7$, $H_{0}=70$ km s$^{-1}$ Mpc$^{-1}$ cosmology, a galaxy
forming all stars within 0.5 Gyr between $z=3$ and $z=2.5$ would
appear as a k+a until $z=1.5$. If it stopped forming stars at $z=2$,
it would still be recognizable as a k+a at $z=1.3$.
The k+a phase of massive ellipticals might be within reach, and possibly
we have begun observing it (van~Dokkum \& Stanford 2003).

\section{Emission-line Spectra and Dust}

The luminosities of some emission lines in an integrated spectrum can 
be used to infer the current SFR in a galaxy because,
as schematically described below for ${\rm H\alpha}$, the luminosity of 
the line is proportional to the number of ionizing photons. In a
star-forming galaxy the number of ionizing photons 
is proportional to the number of massive young stars
and, for a given stellar initial mass function, to the total current SFR:

\begin{equation}
L_{\rm H\alpha} \propto N_{ion-photons} \propto N_{massive-stars} \propto
SFR.
\end{equation}

In distant galaxy studies,
the [O~{\sc ii}] $\lambda$3727 line is often used instead of ${\rm H\alpha}$ 
because the latter is redshifted out of the optical window.
Spectrophotometric models are used to determine the proportionality coefficient
between $N_{ion-photons}$ and $N_{massive-stars}$, and they provide
the standard relations between the SFR and the luminosity of the lines 
(Kennicutt 1992, 1998):

\begin{equation}
SFR = 0.9 \times 10^{-41} \, L_{\rm H\alpha} \, E_{\rm H\alpha} \,\,\,\, M_{\odot} \, {\rm yr}^{-1},
\end{equation}

\begin{equation}
SFR = 2.0 \times 10^{-41} \, L_{\rm [O~{\sc II}]} \, E_{\rm H\alpha} \,\,\,\, M_{\odot} \, {\rm yr}^{-1},
\end{equation}

\noindent
where the line luminosity is in $\rm erg \, s^{-1}$
and $E_{\rm H\alpha}$ is the extinction correction factor
at ${\rm H\alpha}$. A major uncertainty in these estimates is, therefore, the
extinction by dust. 

Modeling has focused on starburst and dusty
galaxies, while little theoretical work has been done on
galaxies forming stars in a continuous, regular fashion, 
which are another sizable component present in distant clusters. 
In the following I will not treat the quiescent star-forming galaxies,
but will summarize the modeling results for starburst and dusty galaxies.
  
Several lines of evidence suggest that dust can
play an important role in some 
distant cluster galaxies. In the optical, spectra
with weak to moderate [O~{\sc ii}] emission and unusually
strong, higher-order Balmer lines were noted to represent a nonnegligible
fraction of both cluster and field spectra (Dressler et al. 1999).
It was realized that in the local Universe such spectra are rare among
normal spirals, while they are common in infrared-luminous starburst
galaxies, which are known to have large dust extinction
(Poggianti et al. 1999; Poggianti \& Wu 2000; see also
Liu \& Kennicutt 1995). Hence, 
regardless of the weakness of their emission lines,
these galaxies were suggested to be in a starburst phase
in which the dust extinction works selectively: the youngest
stellar generations are more affected by dust obscuration than older
stellar populations, which have had time to free themselves or drift
away from their dust cocoons. 

Models with selective extinction are empirically motivated by
observations of star-forming regions in nearby galaxies, and
they explain why different values of $E(B-V)$ are usually measured within
the same spectrum when using different spectral features,
for example why the extinction measured from the emission lines
is usually stronger than that measured from the continuum.
Selective extinction is consistent with the fact that in the
nearby Universe galaxies with the highest current SFR are
generally {\it not} those with the strongest emission lines,
which tend to be low-mass, very late-type galaxies of moderate 
total SFR.

Quantitative modeling of these dusty spectra have been done by
Shioya \& Bekki (2000), Shioya et al. 2001, 
Bekki et al. (2001), and Poggianti, Bressan, 
\& Franceschini (2001c).
While a dust-screen model, with any extinction law,
preserves the equivalent width of the lines (because it affects both
the line and the underlying continuum by a proportional amount),
an age-selective extinction can produce this
peculiar spectral combination with weak emission lines (originating in regions 
with highly extincted, young massive stars)
and exceptionally strong, higher-order Balmer absorption lines
(from exposed intermediate-age stars).

While a spectrum with moderate emission and unusually strong Balmer lines
in absorption is therefore a good candidate for a starburst galaxy,
it has been shown that this modeling is highly degenerate, that this
peculiar type of spectra does not guarantee SFRs above a certain
limit, and that the total SFR
remains unknown without dust-free SFR
estimators such as far-infrared or radio continuum fluxes. 
Radio continuum observations of a cluster at $z=0.4$, for example, 
surprisingly have detected some of the strongest (i.e. youngest)
k+a galaxies in the cluster (Smail et al. 1999),
suggesting that star formation might 
still be ongoing and be totally obscured at [O~{\sc ii}] 
by dust 
in some of the k+a's (see also the $\rm H\alpha$ detection
in some k+a's by Balogh \& Morris 2000 and Miller \& Owen 2002).
Mid-infrared 15 $\mu$m data obtained with ISOCAM have been used
to estimate the far-infrared flux and, therefore, the
SFR in distant clusters (Duc et al. 2002). 
As described in Duc et al. (2003), ISOCAM
observations of a cluster at $z=0.55$ have yielded an unexpectedly high
number of cluster members that are luminous infrared galaxies
with exceptionally high SFRs.

\section{Passive Galaxies and Evolutionary Links}

A large fraction of galaxies in clusters up to $z=1$ have passive
(non-k+a, non-emission-line) spectra.
The evolutionary histories of the passive galaxies are the subject
of other reviews and contributed talks in these proceedings
(see Franx 2003 and Treu 2003); thus, I will limit the following
discussion to a description of the most fundamental contributions 
given to this area of research by spectrophotometric models.

The ages of the stellar populations in luminous, early-type galaxies in 
clusters are known to be old.  Solid evidence for this comes from the 
analysis and evolution of the red color-magnitude (CM) sequence in clusters, 
whose slope, scatter, and zeropoint indicate a passive evolution of stars 
formed at $z>2-3$ (Bower, Lucey, \& Ellis 1992; Arag\'on-Salamanca, Ellis, \& 
Sharples 1993; Rakos \& Schombert 1995; Stanford, Eisenhardt, \& Dickinson 
1995, 1998; Stanford et al. 1997; Schade et al. 1996; Ellis et al. 1997; 
Schade, Barrientos, \& Lopez-Cruz 1997; Barger et al. 1998; Gladders et al. 
1998; Kodama et al.  1998; van~Dokkum et al. 1998, 1999, 2000; De Propris
et al. 1999; Terlevich et al. 1999; Terlevich, Caldwell, \& Bower 2001; 
van~Dokkum \& Franx 2001; Vazdekis et al. 2001). 
Studies of the fundamental plane, mass-to-light ratios, and
the magnesium-velocity dispersion relation agree with these findings
(Bender, Ziegler, \& Bruzual 1996; van~Dokkum \& Franx 1996; Kelson et al. 
1997, 2000, 2001; Ziegler \& Bender 1997; Bender et al. 1998; van~Dokkum et 
al. 1998; Ziegler et al. 2001), though they are necessarily limited
to the brightest subset of galaxies.

\begin{figure*}[t]
\includegraphics[width=1.00\columnwidth,angle=0,clip]{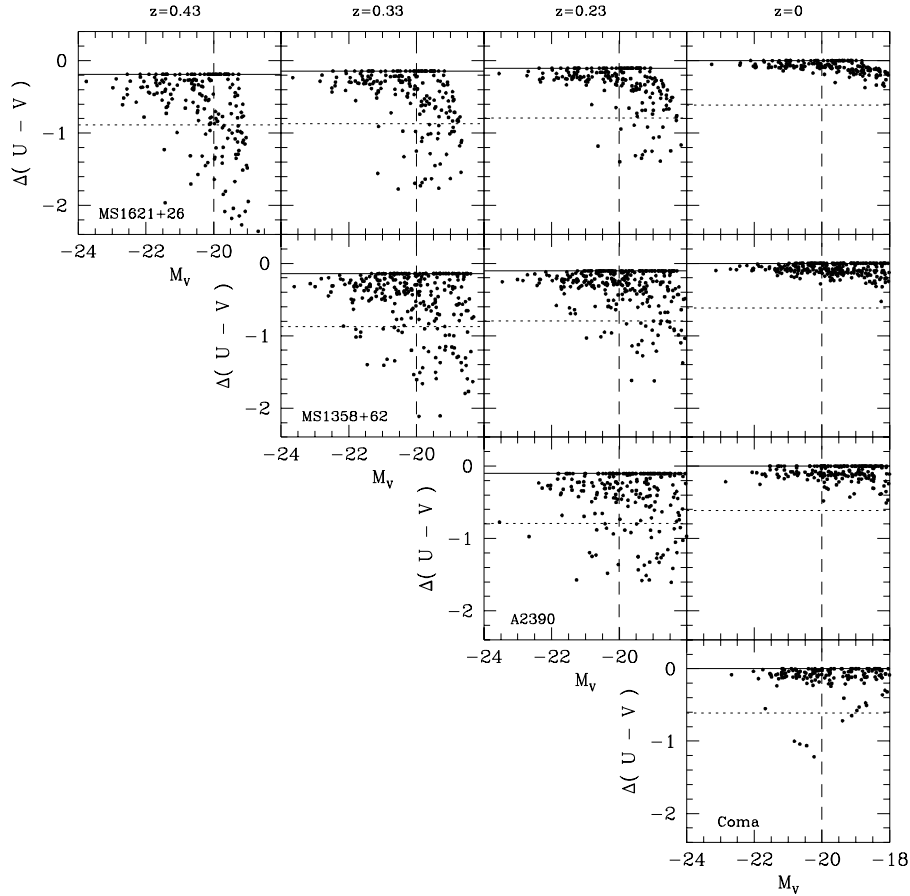}
\vskip 0pt \caption{
Evolution of the CM diagram (Kodama
\& Bower 2001). From left to right, redshift is decreasing as shown
on the top of the plots. The leftmost panels are observed CM 
diagrams of clusters, while the subsequent evolution are examples of 
Monte-Carlo simulations in which star formation is truncated after accretion
of galaxies from the field. The CM diagram of Coma is shown for comparison
in the bottom right panel.
\label{fig:cm_faded_decay.ps}}
\end{figure*}

How can these results be reconciled with the presence of numerous blue
galaxies in distant clusters (the Butcher-Oemler effect), given
that these blue, star-forming galaxies have largely ``disappeared''
(i.e. become red) by $z=0$? This has been investigated by works
that have modeled the evolution of galaxy colors and magnitudes
(Bower, Kodama, \& Terlevich 1998;
Smail et al. 1998; Kodama \& Bower 2001). Figure~\ref{fig:cm_faded_decay.ps}
 presents the results
of Kodama \& Bower (2001), who have shown that by evolving the observed
CM diagram of intermediate-redshift clusters it is possible
to obtain a diagram at $z=0$ that is mostly composed of red galaxies,
similar to the observed CM diagram of the Coma cluster.
Moreover, it is now clear that the CM red sequence of clusters is 
comprised not only of early-type galaxies, but also contains
morphologically classified
spirals (Poggianti et al. 1999; Couch et al. 2001; Terlevich et al. 2001;
Balogh et al. 2002; Goto et al. 2003a, b).

These results imply that the CM sequence today is composed of
a varied population of galaxies with different star formation histories.
As shown by several works, including the one by Kodama \& Bower (2001),
although the principal driver of the CM sequence is the
correlation between galaxy luminosity and metallicity, there is still
room for a relatively recent epoch of star formation activity
in a significant fraction of the (today) red galaxies.
When contrasting this with the homogeneity and old ages of 
red sequence galaxies derived from CM and fundamental plane
studies, it is important to 
take into account two aspects: the morphological transformations
in clusters and the evolution of the galaxy luminosities.

\subsection{Morphological Evolution}

The morphological mix of galaxies in
clusters at various redshifts strongly suggests
that a significant fraction of the spirals in distant clusters have evolved
into the S0s or, more generally, the early-type galaxies that dominate
clusters today (Dressler et al. 1997; Fasano et al. 2000; van~Dokkum et al. 
2000; Lubin, Oke, \& Postman 2002; see Dressler 2003).
This morphological evolution is likely to lead to a ``progenitor bias'':
in distant clusters we would be observing as early-type galaxies only
the ``oldest'' subset of the present-day early-types, those that were already
assembled and stopped forming stars at high redshift
(van~Dokkum \& Franx 1996, 2001; Stanford et al. 1998).

\begin{figure*}[t]
\includegraphics[width=1.0\columnwidth,angle=0,clip]{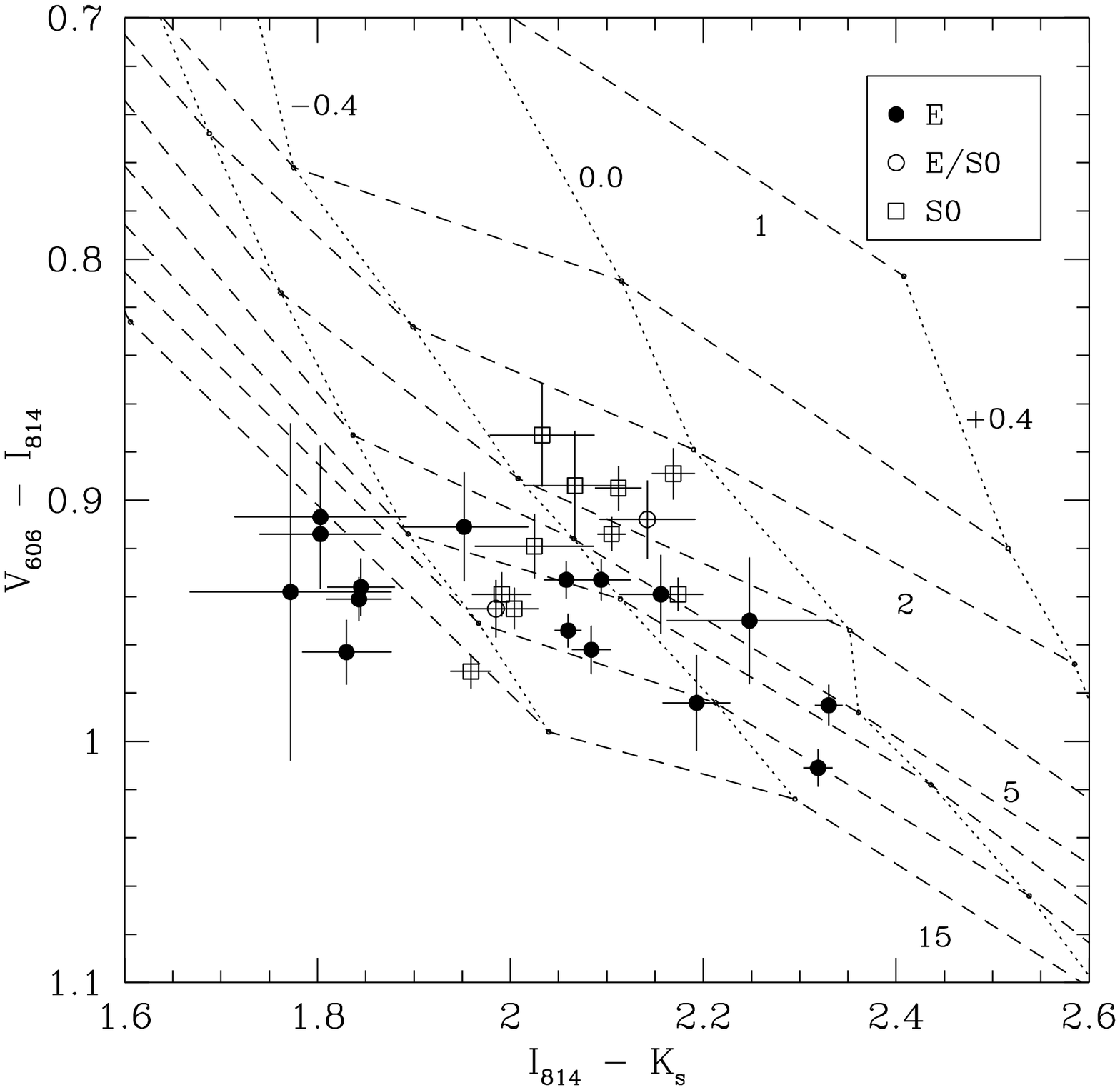}
\vskip 0pt \caption{
Color-color diagram of galaxies in Abell 2218 (Smail et al.
2001). The grid of lines shows models of single stellar populations
of constant age (pseudo-horizontal lines) and constant metallicity 
(pseudo-vertical lines). As shown in the legend of the plot, the 
group of galaxies with younger ages (lower $V-I$ colors)
than the rest are all S0 galaxies.
\label{fig:fig5.ps}}
\end{figure*}

The signature of relatively recent star formation activity in some
of the early-type galaxies in clusters has been searched for in several ways.
Recently, significant differences between the ages of the stellar populations
of ellipticals and a fraction of the
S0 galaxies have been detected, supporting the 
scenario of spirals evolving into S0s
(Kuntschner \& Davies 1998; van~Dokkum et al. 1998;
Terlevich et al. 1999; Poggianti et al. 2001b; Smail et al. 2001; Thomas 2002).
Many of these works derive luminosity-weighted ages and metallicities
by comparing a metallicity-sensitive and an age-sensitive
spectral index with a grid of spectrophotometric models.
Similarly, an age-sensitive and a metallicity-sensitive {\it color
index} can be used, as shown in Figure~\ref{fig:fig5.ps} (Smail et al. 2001).
The plot shows the color-color diagram of galaxies in Abell 2218 at $z=0.17$,
where a group of faint S0 galaxies is seen to lie at younger
luminosity-weighted ages (lower $V-I$ color) than all the ellipticals and
the rest of the S0s.
The possibility of a spiral-to-S0 transformation has been quantitatively
investigated in terms of galaxy numbers and morphologies
by Kodama \& Smail (2001) 
and from the spectrophotometric point of view by Bicker, Fritz-v.~Alvensleben,
\& Fricke (2002),
providing interesting constraints on the possible evolutionary paths.

However, not all studies find differences between S0s and ellipticals.
Ellis et al. (1997), J{\o}rgensen (1999, and references therein), 
Jones, Smail, \& Couch (2000), and Ziegler
et al. (2001) do not detect a significant morphological dependence
of the stellar population ages and metallicities in early-type
cluster galaxies. It is important to stress that the presence of recent
star formation in S0s has never been detected in the {\sl brightest}
S0s but becomes a prominent effect when probing fainter down the
luminosity function. For example, Coma S0s with recent star
formation are fainter than $M^{\star} +1.3$, as expected given the
luminosities of their possible spiral progenitors at intermediate
redshift (Poggianti et al. 2001b). When studying the dependence of
the stellar population properties on the morphological type of galaxies,
it is therefore important to simultaneously disentangle the
dependence on the galaxy luminosity and to consider the effects of 
luminosity evolution. This is further discussed in the following
section.

\subsection{Luminosity Evolution}

\begin{figure*}[t]
\includegraphics[width=1.0\columnwidth,angle=0,clip]{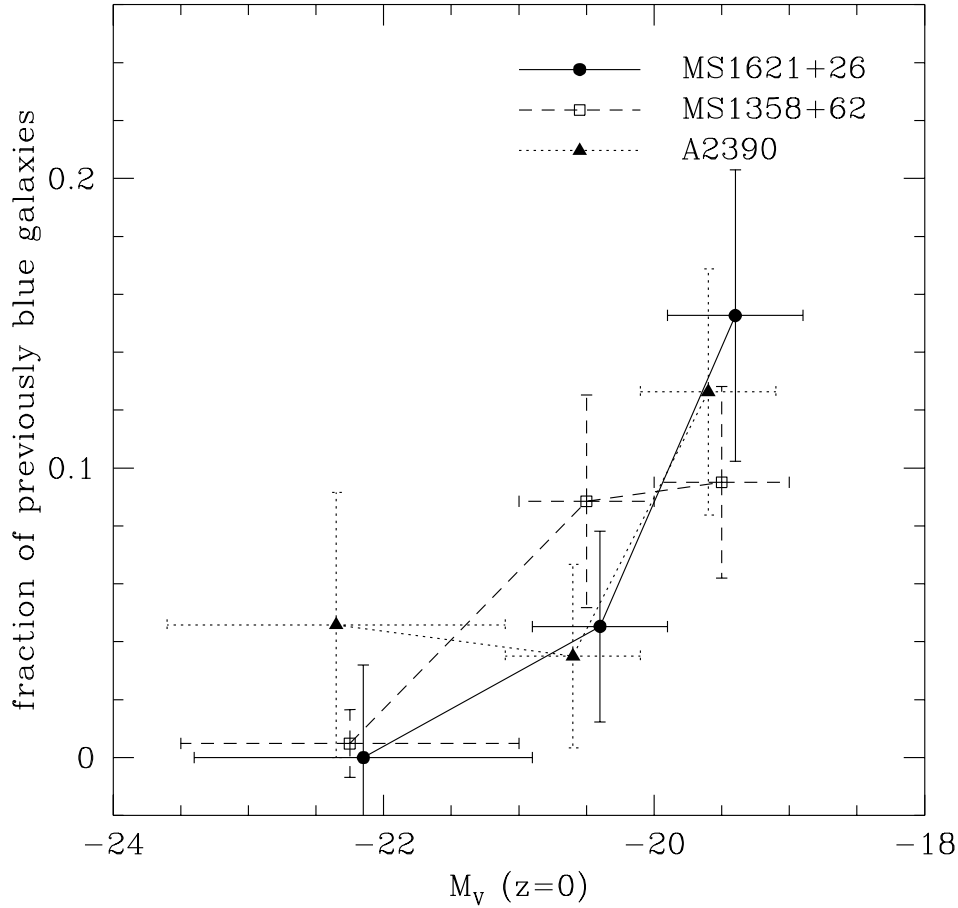}
\vskip 0pt \caption{
Fraction of galaxies observed as blue at intermediate
redshifts (three clusters at $z=0.2$, 0.3, and 0.4)  
as a function of the present-day
($z=0$) absolute $V$ magnitude. The $z=0$ magnitude is computed
assuming a decline of star formation as a consequence of
the accretion onto the cluster. (From Kodama \& Bower 2001.)
\label{fig:downsizing.ps}}
\end{figure*}

There are two aspects of the evolution of galaxy luminosities
that are important to stress. The first aspect, already mentioned 
above, is the {\sl fading}. As discussed in \S 1.2,
the interruption of star formation seems to be a phenomenon 
involving a significant fraction of the cluster galaxies.
As a consequence, galaxies not only become redder but also fainter:
this evolution in luminosity is obviously stronger than the evolution
occurring both in the case of passive evolution of an old population
and in the case of a galaxy continuously forming stars with no
large variation in SFR until $z=0$. 
Kodama \& Bower (2001) computed the expected fraction of previously
blue galaxies as a function of the present-day absolute magnitude,
as shown in 
Figure~\ref{fig:downsizing.ps}, starting from the CM diagram 
of three intermediate-redshift clusters and computing the luminosity
evolution using models with truncated star formation. 
The figure shows how the fraction of present-day
red galaxies that were blue at higher redshift increases progressively
as one goes fainter. The relevance of the fading and the evolution 
as a function of galaxy luminosity are discussed also in many other 
works (Postman, Lubin, \& Oke 1998; Smail et al.  1998, 2001;  van~Dokkum 
et al. 1998; Terlevich et al. 1999; Ferreras \& Silk 2000; Poggianti et al. 
2001b; Shioya et al. 2002; Merluzzi et al. 2003; Tran et al. 2003).

Related to this, there is a second interesting point related to luminosity,
known as the ``downsizing effect'': the fact that, as one goes to lower
redshifts, the maximum luminosity/mass of galaxies with significant star
formation activity seems to be progressively decreasing, possibly both in
clusters (Bower et al. 1999) and in the field (Cowie et al. 1996).

An attempt to derive the distribution of luminosity-weighted ages
as a function of galaxy luminosity over a range of
almost 7 magnitudes has been done for Coma in a new magnitude
limited spectroscopic survey (Poggianti et al. 2001a). In this study 
spectrophotometric models were also used 
to derive luminosity-weighted
ages for spectra without emission lines (the great majority in Coma)
from index-index diagrams.
These ages give a rough estimate of the time elapsed since the latest
star formation activity stopped. This time has been transformed into
a cosmological epoch (i.e. redshift) adopting an $\Omega_{\Lambda}=0.7$,
$H_{0}=70$ km s$^{-1}$ Mpc$^{-1}$ cosmology. About half of all 
non-emission-line galaxies of any magnitude in Coma
show no detectable star formation activity during the last 9 Gyr, i.e.
since $z \approx 1.5$. The other half 
{\sl do} show signs of some star formation below this redshift and, interestingly, display
a trend of epoch of latest star formation with 
galaxy magnitude: recent ($z<0.25$) star formation is detected
in 25\%--30\% of the dwarf galaxies, while only in 5\%--10\% of the giant
galaxies. In contrast, 30\%--40\% of the giants
reveal some star formation at intermediate redshifts ($0.25 <z < 1.5$),
compared to 15\%--20\% for the dwarfs.

\begin{figure*}[t]
\includegraphics[width=1.0\columnwidth,angle=0,clip]{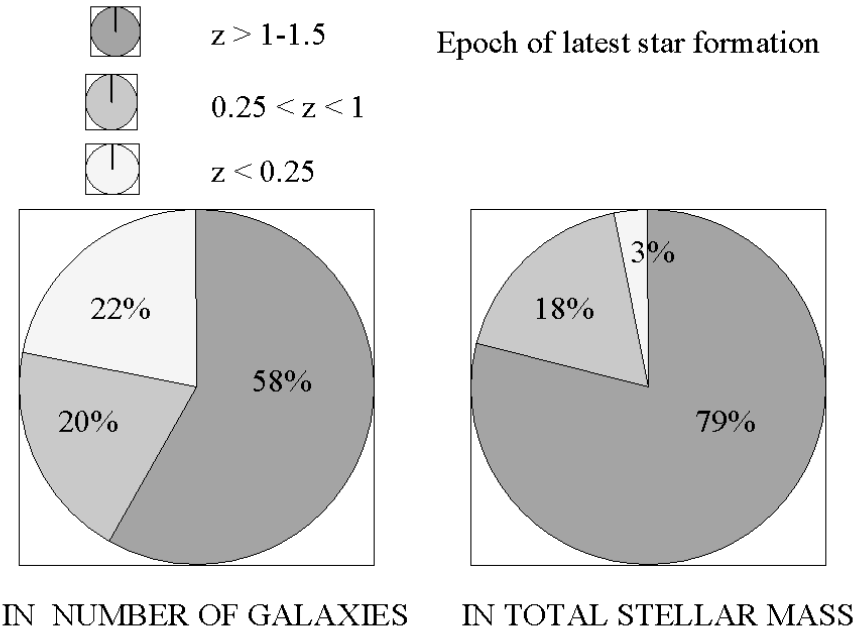}
\vskip 0pt \caption{
{\it Left}: Fractions of non-emission-line Coma galaxies with signs of 
star formation in three cosmological intervals. These have been computed
from the luminosity-weighted ages derived from spectral index diagrams
in Poggianti et al. (2001a). {\it Right}: The number of galaxies
have been transformed into fractions of stellar mass in galaxies
with signs of star formation during the three epochs (see text).
\label{fig:solo.PS}}
\end{figure*}

Overall, including all galaxies with $M_B$ between $-21$ and $-14$ mag, 
the fraction of galaxies with signs of star formation 
at the various epochs is presented in Figure~\ref{fig:solo.PS}  (left).
Going a step forward, one can try to quantify the stellar mass
involved, transforming the observed luminosities into a mass by
adopting the mass-to-light ratio of a single stellar population
with an age equal to the luminosity-weighted age derived from the
spectral indices. As shown in the right panel of Figure~\ref{fig:solo.PS},
galaxies with some star formation below $z=1$ account for
only about 20\% of the mass, and those with some star formation at $z<0.25$
for only 3\% of the mass, being mostly low-mass dwarf galaxies.
Note that this is {\sl not} the mass fraction formed at each epoch,
but it is the {\sl total} mass in those galaxies that had
some (unquantified) star formation.
An example of an attempt to estimate the total mass fraction
formed at each redshift using population synthesis models of cluster
galaxy colors and magnitudes
can be found in Merluzzi et al. (2003), who find, for quiescent,
non-starbursting star formation histories, that about 20\% of the mass was 
formed below redshift 1.

Other evidence of a downsizing effect comes from k+a galaxies
in clusters at low redshift. A number of works have investigated
the presence of Balmer-strong absorption-line galaxies in nearby
clusters, especially in Coma (Caldwell et al. 1993, 1996; Caldwell \& Rose 
1997, 1998; Caldwell, Rose, \& Dendy 1999;  Castander et al. 2001;
Rose et al. 2001). Some of these works 
have highlighted the tendency of these Balmer-strong galaxies to be
intrinsically fainter than the k+a galaxies observed
at high redshift. In our Coma survey (Mobasher et al. 2001; Poggianti
et al. 2001a) we have found no luminous k+a galaxy down to a
magnitude-limit comparable to the MORPHS limit at $z=0.5$, while
there are strong and frequent cases among faint Coma galaxies
at $M_V> -18.5$ mag. At these magnitudes 10\% to 15\% of the cluster
galaxies are k+a's, and many of these dwarf spectra are exceptionally strong, 
post-starburst galaxies (Poggianti et al. 2003).

\section{Summary}

In this contribution I have summarized 
the main achievements of spectrophotometric modeling
of galaxy evolution in distant clusters. Five main points have
been mentioned:

\begin{enumerate}

\item
k+a spectra unequivocally indicate post-star-forming or post-starburst
galaxies, depending on the strength of the Balmer lines.

\item
A galaxy with weak to moderate emission lines {\it and} strong
Balmer lines in absorption is a good candidate for an ongoing starburst,
or a galaxy currently forming stars at a vigorous rate.

\item
There are observed differences in the 
stellar population ages between ellipticals and S0 galaxies
that are consistent with a significant fraction of the latter 
having evolved from star-forming spirals at intermediate redshifts.
This result is not universal, though, and probing sufficiently faint
magnitudes seems to be crucial for detecting any difference, at least
at low redshift.

\item
The CM sequence in clusters today appears to be a mixed bag.
The slope is dominated by metallicity, but there is room for
relatively recent star formation in a significant fraction of the galaxies.
The bright end of the CM sequence seems to be populated
by homogeneously old galaxies, while more and more remnants of galaxies
with star formation at intermediate redshifts ($z \approx 0.5$)
can be found as one goes fainter along the sequence.

\item
Related to the previous point, a ``downsizing effect'' is observed in
clusters, probably in parallel to a similar effect in the field:
toward lower redshifts, the maximum luminosity/mass of galaxies with
significant star formation activity progressively decreases. The
downsizing effect could be responsible of the fact that k+a cluster
spectra appear to occur in luminous galaxies at $z=0.5$ and in faint
(mostly dwarf) galaxies at $z=0$.

\end{enumerate}

In the next few years the efforts of 
spectrophotometric modeling will likely
concentrate on modeling at higher redshifts than it has been done so far, 
in order to aid the interpretation of the large photometric and 
spectroscopic data sets that are being acquired in clusters at $0.5 < z< 1.5$.
We can expect further progress also at low redshift, where 
the model improvements, especially regarding the metal abundance ratios,
coupled with the new large surveys of nearby clusters and groups,
are expected to reveal detailed characteristics
of cluster galaxies that can be helpful in constraining
their formation histories. As more and better data are obtained,
the most important contribution of the modeling will be 
the endeavor to reach a quantitative general picture that can
simultaneously account for the observations at all redshifts.

\vspace{0.3cm}
{\bf Acknowledgments}.
The author wishes to thank the organizers for a very pleasant and
interesting Symposium, A. Oemler for reviewing this paper and
suggesting numerous improvements to the text, and L. Ho for careful and
patient copy-editing.

\end{document}